\documentclass[aps,prd,11pt,onecolumn,superscriptaddress,showpacs, nofootinbib]{revtex4-1}
\pdfoutput=1
\usepackage{hyperref}
\usepackage{graphicx}
\usepackage{enumerate}
\usepackage{amsfonts,amsmath,amssymb,bm,bbm}
\usepackage{color}
\usepackage{epsfig, subfigure}
\usepackage{setspace}
\usepackage{booktabs, tabularx} 

\hypersetup{
    pdfnewwindow=true,      
    colorlinks=true,       
    linkcolor=blue,          
    citecolor=blue,        
    filecolor=blue,      
    urlcolor=blue        
}

\newcommand{\be}{\begin{equation}}  
\newcommand{\ee}{\end{equation}}
\newcommand{\bea}{\begin{eqnarray}}  
\newcommand{\eea}{\end{eqnarray}}

\let\oldsqrt\sqrt
\def\sqrt{\mathpalette\DHLhksqrt}
\def\DHLhksqrt#1#2{%
\setbox0=\hbox{$#1\oldsqrt{#2\,}$}\dimen0=\ht0
\advance\dimen0-0.2\ht0
\setbox2=\hbox{\vrule height\ht0 depth -\dimen0}%
{\box0\lower0.4pt\box2}}

\begin{document}

\title{\mbox{Leptonic CP Violation Phases, Quark-Lepton Similarity and Seesaw Mechanism}}

\author{Basudeb Dasgupta}
\email{bdasgupta@ictp.it}

\affiliation{{International Center for Theoretical Physics, 
Strada Costiera 11, 34014 Trieste, Italy.}}

\author{Alexei Yu. Smirnov}
\email{smirnov@mpi-hd.mpg.de}
\affiliation{{International Center for Theoretical Physics, Strada Costiera 11, 
34014 Trieste, Italy.}}
\affiliation{{
Max-Planck-Institute for Nuclear Physics,
Saupfercheckweg 1, D-69117 Heidelberg, Germany.}}

\date{\today}

\begin{abstract}
We explore  generic features of  the leptonic CP violation in the framework 
of the seesaw \mbox{type I} mechanism with similarity of the  Dirac lepton  and quarks mass 
matrices $m_D$. 
For this, we elaborate on the standard parametrization conditions  
which allow to simultaneously obtain the Dirac and Majorana phases. 
If the only origin of CP violation is the left-handed (LH) transformation 
which diagonalizes $m_D$  (similar to  quarks), the 
leptonic CP violation is suppressed and  
the Dirac phase is close to $\pi$  or to $0$ 
with  $\sin \delta_{CP} \approx  (\sin \theta_{13}^q /\sin \theta_{13}) \cos \theta_{23} 
\sin \delta_q   \sim   \lambda^2 \sin \delta_q$. Here  
$\lambda \sim \theta_C$, is the Cabibbo mixing angle, 
and $\theta_{13}^q$ and $\theta_{13}$ are the 1-3 mixing angles of  quarks and leptons
respectively. The Majorana phases $\beta_1$ and $\beta_2$ are
suppressed as $\lambda^3\sin\delta_q$. 
For Majorana neutrinos implied by seesaw,  
the right-handed (RH) transformations are important. We explore the simplest extension inspired 
by Left-Right (L-R) symmetry with small CKM-type CP violation. 
In this case,  seesaw enhancement of the CP violation   
occurs due to strong hierarchy of the 
eigenvalues of $m_D$ leading to $\delta_{CP} \sim 1$. 
The enhancement is absent under the phase factorization conditions which require 
certain relations between  parameters of the Majorana mass matrix of RH neutrinos.

\end{abstract}


\maketitle

\section{Introduction}
\label{sec:Intro}

Establishing the leptonic CP violation is one of the major experimental 
frontiers in neutrino physics. 
The Dirac and Majorana CP phases are among the few yet unknown parameters for which 
a prediction may still be made.
So, we need  
to understand what  particular values or intervals of the CP phases will imply 
for fundamental theory.

Indeed, there are numerous predictions of the phases which are  
based on broad spectrum of ideas, approaches, and models~\cite{reviews}. 
Some approaches that have been employed are 
(i) Neutrino and charged lepton mass matrices with certain properties such as -- textures
\cite{structure}, symmetries, and symmetry violations, 
e.g., $\mu - \tau$ reflection or generalized symmetry \cite{mutau};   
(ii) Models with discrete flavor symmetries~\cite{discrCP}, 
which can  realize geometric origins of the phases,  
the  CP violation due to group structure or complex Clebsch-Gordan coefficients \cite{geom},
or connect the phases and the mixing  angles \cite{dani1}, {etc.};  
(iii) Grand unification with seesaw type I and  type II \cite{rabi}
\footnote{The extreme possibility is that the mixing of quarks and leptons concides at the GUT scale and 
the low energy difference is due to large  renormalization 
group evolution for a quasi-degenerate mass spectrum \cite{equal}.};  
(vi) Radiative  generation~ of CP violation~\cite{radiative};
(vii) Relating the leptonic CP phase to other physics, e.g., a solution to the strong CP problem 
wherein $\delta_{CP}=0$ or $\pi$ is predicted~\cite{kuchimanchi}.
Many efforts have been devoted to obtain maximal CP violation, i.e., $\delta_{CP} = \pi/2$ 
\cite{max-cp}, although other values essentially from 0 to $\pi$ have also been found. 

Can we really  predict the leptonic CP phase, given 
that even in the quark sector, where all parameters are known,  
there is no unique and convincing explanation 
of the value of CP phase?  Moreover, in the lepton sector the situation is expected to be 
more complicated due to presence of additional structures 
which are responsible for the smallness of neutrino masses.  
Can the lepton and quark CP phases be equal, or connected in some way? 
To address these questions it is instructive to represent the lepton mixing matrix 
in the form 
\be 
U_{PMNS} =  U_L U_X,  
\label{product}
\ee
where $U_{L}$ is somehow related to the quark CKM-mixing matrix and  $U_X$ 
reflects new physics responsible for smallness of neutrino mass 
and large mixing angles \cite{tanimoto}, 
\cite{qlc}, \cite{xing1}, \cite{harada}, \cite{antusch}, \cite{yasaman},   
\cite{picariel}.  Here  $U_L$ and  $U_X$ 
can follow from diagonalization of mass matrices of the charged leptons,  $U_L = U_l^{\dagger}$,  
and neutrinos, $U_X = U_\nu$, respectively.   
Origins of CP violation can be in  $U_l$ \cite{Acosta:2014dqa} 
and/or $U_\nu$. 
The assumption  $U_l \sim V_{CKM}$ corresponds to 
the Quark-Lepton Complementarity \cite{qlc}, so that   
$U_{PMNS} = V_{CKM}^{\dagger} U_X$.  
This possibility has been explored for $U_X = U_{BM}$ (bimaximal mixing matrix)~\cite{xing1} and 
$U_X = U_{TBM}$ (tribimaximal mixing matrix) referred as Cabibbo-TBM \cite{cab-tbm}. 
In these cases the origin of CP could be in $V_{CKM}$ or in  
the diagonal phase matrix  attached to  $U_X$. 
In \cite{xing1}  the ``correlation matrix'' $U_X$ 
has been taken in the form $U_X  = P (\phi_l) U_{BM} Q(\phi_i)$,  
where  $P(\phi_l)$ and $Q(\phi_i)$ are diagonal phase matrices.  
It was noticed that if $\phi_l = 0$, the  Jarlskog invariant  
is very small~\cite{xing1}, \cite{yasaman}, \cite{harada}:  
$J_{PMNS} = \sin \theta_{13}^q \sin \delta_{CKM}$, { i.e.}, too small to be 
measured in future experiments. 

The ansatz (\ref{product}) can be naturally realized in the seesaw type I mechanism ~\cite{seesaw}
which is the simplest and the most natural explanation of smallness of 
neutrino masses  
as well as large lepton  mixing~
\cite{ss-enhance}. 
It is simplest because  only RH  
neutrinos are added to the theory. 
It is natural in the sense that it allows to explain 
smallness of neutrino mass and the substantial difference between 
lepton mixing and quark mixing, while at the same time maximally 
implementing the  quark-lepton  similarity. The latter, in turn, 
is expected, e.g., in Grand unified theories.  
Seesaw type I mechanism with similar Dirac 
mass matrices for neutrinos and  quarks  
defines the canonical seesaw mechanism. 

In this paper we consider the  leptonic CP phases that can arise from 
this canonical seesaw mechanism, 
which provides the closest possible connection of the quark and lepton sectors. 
We will further generalize the  relation (\ref{product}) 
assuming that $U_L$ 
has similar to  $V_{CKM}^{\dagger}$ structure   
but in general does not coincide with $V_{CKM}^{\dagger}$. 
For the matrix $U_X$  we will not assume any special structure 
but restrict it only  by the condition that 
the product (\ref{product}) reproduces the 
experimentally observed  values of the mixing angles. 
We will find the phases in the standard parametrization of the PMNS matrix. For this 
we formulate and use the standard parametrization conditions which 
allow us to obtain simultaneously both the Dirac  
and Majorana CP phases.

We  find that if the only source of CP violation is the Kobayashi-Maskawa (KM) -type phase 
in $U_L$,  it leads to a small  $\delta_{CP}$. In the seesaw 
mechanism due to the Majorana nature of neutrinos
the CP violation in the RH sector become relevant for the PMNS CP phases. 
That includes the phases 
in the RH rotation $U_R$ that diagonalizes the Dirac mass matrix $m_D$ 
as well as in the Majorana mass matrix of RH neutrinos, $M_R$. 
We find that generically  
the seesaw mechanism  enhances CP violation that appears in $U_R$,  so  that 
$\delta_{CP} = {\cal O} (1)$.  Such an enhancement is absent and 
the CP phases are small (or close to $\pi$) 
if  parameters of $M_R$ satisfy certain (phase factorization) relations.  
We find relations between the Dirac and Majorana phases which can be used 
to test these scenarios.
An observation of (large) leptonic CP violation in oscillation experiments and/or neutrinoless 
double beta decay would therefore be a signature that there is a new source of CP violation, 
beyond the leptonic analogue of KM-phase and coming from the RH sector, 
or that neutrino masses do not arise from a canonical seesaw. 

We will argue that specific values of  the leptonic CP phases are  possible 
since the  contribution  of the Kobayashi-Maskawa type phase turns out to be suppressed 
or be close  to $\delta_{CP} \sim \pi$ and the  
main contribution comes from the RH sector which can  obey 
specific symmetries.

The paper is organized as follows: In Sec.\,\ref{sec:Formalism} we present the formalism 
for calculating the CP phases. In Sec.\,\ref{sec:Dirac} we derive the expressions for the CP phases 
when the only source of CP violation is a KM-like phase in the left-handed transformation that diagonalizes 
the Dirac mass matrix.  
The general case with CP violation in the RH sector is considered in Sec.\,\ref{sec:general}.  
In Sec.\,\ref{sec:UR}  we explore a specific case of CP violation in the RH sector, 
which corresponds approximately to a L-R symmetry of the theory.  
In Sec.\,\ref{sec:factorization} we consider special 
conditions where the resulting CP phase is still small.  
We then show that in general the seesaw enhancement of CP violation occurs which leads to 
$\delta_{CP} \sim {\cal O} (1)$,  even  if CP violation in $U_R$ is of KM-type.  
We present some phenomenological consequences in Sec.\,\ref{sec:consequences} and 
conclude in Sec.\,\ref{sec:conclude}.

\section{Seesaw Type I, CP violation, and Standard Parametrization}
\label{sec:Formalism}

\subsection{$U_X$ matrix in seesaw type I}
\label{sec:UXinseesaw}

We introduce the Dirac matrix, $m_D$, in the flavor basis  and Majorana mass matrix, $M_R$, 
according to the mass terms of the Lagrangian
\begin{equation}
{\cal L}_{mass}=-\bar{\nu}_L m_D \nu_R -  \frac{1}{2} \nu_R^T C^\dagger M_R \nu_R\, + h.c.\, .  
\nonumber
\end{equation}
Integrating out the RH neutrinos we obtain 
${\cal L}_{mass}=-\bar{\nu}_L m_{\nu} \bar{\nu}_L^T + h.c. $,   where 
the matrix of light neutrinos  
in the flavor basis equals 
\begin{equation}
m_{\nu}= - m_D M_R^{-1} m_D^T\,.
\nonumber
\label{eq:seesaw}
\end{equation}
The Dirac mass matrix
can be represented in the flavor basis  as 
\begin{equation}
m_D = U_{L} m^{diag}_{D} U_{R}^{\dagger}\,,  
\label{eq:dirac}
\end{equation}
where $U_{L}$ and  $U_{R}$ are the transformations 
$\nu_L = U_L \nu_{L}^{diag}$, $\nu_R = U_R \nu_{R}^{diag}$\,,
that diagonalize $m_D$, and $m^{diag}_{D} \equiv {\rm diag} (m_{1D}, m_{2D},  m_{3D})$.   
The light neutrino mass matrix in the flavor basis is   
\begin{equation}
m_{\nu} = U_{PMNS} ~ m_\nu^{diag}~U_{PMNS}^T\,,
\label{eq:mPMNS}
\end{equation}
where  
\begin{equation}
\nu_L = U_{PMNS}\nu_{mass}\,, 
\nonumber
\end{equation}
are the light neutrino flavor states
and $m_\nu^{diag} = {\rm diag} (m_1, m_2, m_3)$ 
is the diagonal matrix of real and positive neutrino mass eigenvalues. 

Inserting (\ref{eq:mPMNS}) and (\ref{eq:dirac}) into the
seesaw  expression (\ref{eq:seesaw}) we obtain   
\begin{equation}
{U}_{PMNS}~{m}^{diag}_\nu~{U}_{PMNS}^T 
= -U_{L} m^{diag}_{D} U_{R}^{\dagger} \frac{1}{M_{R}} U_R^{*} m^{diag}_{D} U_{L}^T \,.
\label{eq:master}
\end{equation}
The relationship in \,(\ref{eq:master}) can be re-expressed as
\begin{equation}
U_{PMNS} {m}^{diag}_\nu {U}_{PMNS}^T = U_L M_X U_L^T\,, 
\label{eq:master2}
\end{equation}
where 
\begin{equation}
M_X \equiv - m^{diag}_{D} U_{R}^{\dagger}\frac{1}{M_{R}} U_R^* m^{diag}_{D}. 
\label{eq:xdef}
\end{equation}
It is the structure of the matrix $M_X$ that  produces the difference in 
masses and mixing of quarks and leptons.

Since $U_{PMNS}$ and $U_L$ are unitary matrices,  the eigenvalues on both sides of 
\,(\ref{eq:master2}) should coincide. 
Therefore $M_X$ can be rewritten as,
\be
M_X = U_X  {m}^{diag}_\nu U_X^{T}, 
\label{eq:crucial}
\ee
and the mixing matrix $U_X$ is  obtained by diagonalization of (\ref{eq:xdef}). 
From  (\ref{eq:master2}) and (\ref{eq:crucial})
we obtain 
\begin{equation}
U_{PMNS} {m}^{diag}_\nu {U}_{PMNS}^T  = 
U_L U_X {m}^{diag}_\nu U_X^T  U_L^T \, , 
\nonumber
\label{eq:mnul} 
\end{equation}
which can be satisfied 
if and only if the matrix $U_L U_X$ coincides  with 
$U_{PMNS}$ up to a diagonal matrix 
$D = {\rm diag}[(-1)^{n}, (-1)^{m}, (-1)^{k}]$, where $n$, $m$, $k$ are integers,   
which is the symmetry transformation of a generic 
diagonal Majorana mass matrix. 
Therefore,  
$U_{PMNS} =  U_L U_X D$. In what follows, we will 
absorb $D$ into the phase matrix of $U_X$. 

Thus, within the seesaw paradigm we arrive at the  relationship 
(\ref{product}) with $U_X$ being the 
matrix which diagonalizes $M_X$ (\ref{eq:xdef}).  
Notice that $U_L$ would be the lepton mixing matrix, 
if $M_X$ is diagonal or there are no Majorana mass terms. 
Whereas,  $U_X$ encodes 
information about the eigenstates of the Dirac and Majorana mass matrices, 
as well as  about mismatch of the $\nu_R$ transformations which  diagonalize 
$m_D$ and $M_R$. The matrix $M_R$ can be written as $M_R = U_M M_R^{diag} U_M^T$, 
so that  $M_R^{-1} = U_M^* (M_R^{diag})^{-1} U_M^{\dagger}$, 
and consequently,  $M_X = - m_D^{diag}  U_R^{\dagger} U_M^*  (M_R^{diag})^{-1} U_M^{\dagger} U_R^* m_D^{diag}$. 
If $U_M = U_R^*$, then according to \,(\ref{eq:xdef}) $M_X$ is diagonal.

In what follows we will explore the relationship expressed in (\ref{product}) to 
derive predictions for the physical CP violating phases in $U_{PMNS}$  in terms of 
the relevant parameters of the RH sector and $U_L$. Results of this section are general and 
can be applied to any mechanism which reproduces~(\ref{product}).

\subsection{Standard parametrization conditions}
\label{sec:SPconds}

Motivated by its widespread use, 
we will consider the CP phases that appear 
in the standard parametrization of the PMNS matrix 
$U_{PMNS}^{std}$~\cite{StandardP}: 
$$
U_{PMNS}^{std}= R_{23}\Gamma_{\delta}R_{13}\Gamma_{\delta}^\dagger R_{12}\,,
$$ 
where $\Gamma_{\delta}\equiv {\rm diag}(1,1,e^{i\delta_{CP}})$  and $\delta_{CP}$  
is the Dirac CP violating  phase. 
Usually to find the CP phase one computes the Jarlskog invariant 
of $U_{li}$, and uses the mixing parameters in the standard parametrization. 
We find that a more instructive and transparent way 
to find CP phases is to compute the  mixing matrix directly   
and reduce it to the standard parametrization form by rephasing.

In general, the PMNS matrix 
can be written as 
\be
U_{PMNS} = D (\phi) U_{PMNS}^{std}(\delta_{CP}) 
\Gamma_M (\beta). 
\label{pmns-gen}
\ee 
Here $D (\phi) \equiv {\rm diag} (e^{i\phi_e}, e^{i\phi_\mu}, e^{i\phi_\tau})$  is 
the matrix of phases which can be eventually absorbed into the wave functions of 
charged leptons, and 
\be
\Gamma_M \equiv {\rm diag}(e^{i\beta_1}, e^{i\beta_2}, 1)   
\nonumber
\ee
is the matrix of the Majorana phases \footnote{An alternate parametrization of $\Gamma_M$ is 
${\rm diag}(1, e^{i\frac{\alpha_{21}}{2}}, e^{i\frac{\alpha_{31}}{2}})$, 
and we can recover these Majorana phases, 
$\alpha_{21}=2(\beta_2-\beta_1)$ and $\alpha_{31}=-2\beta_1$, 
by an overall rephasing of $U_{PMNS}$ from the right side by $e^{-i\beta_1}$.}. 
We will use the standard parametrization also for  
the  matrices on the RH side of ~(\ref{product}):  
\be
U_L  = D (\psi) U_L^{std} (\delta_L) D(\chi), ~~~  
U_X =  D (y) U_X^{std} (\delta_X) D(z),  
\label{eq:ulux}
\ee
where $D({ \psi}) \equiv {\rm diag} (e^{i \psi_e}, e^{i \psi_\mu}, e^{i \psi_\tau})$, 
{etc.},    
$U_L^{std}$ and $U_X^{std}$ are the matrices in the standard 
parametrizations which contain a single CP phase each. 
Then the product of matrices in Eq. (\ref{product})  equals 
\begin{equation}
U_L  U_X =  D(\psi) U_L^{std}(\delta_q) 
D (\alpha) U_X^{std} (\delta_X) D(z), ~~{\rm where}~~\alpha_l \equiv \chi + y. 
\label{eq:ulux1}
\end{equation}
Clearly, introduction of the two separate matrices  $D (\chi) $ and  $D(y)$ 
is irrelevant for the  light neutrino mass matrix but   
it does matter for the structure of the RH sector. 

Inserting expressions  (\ref{pmns-gen}) and (\ref{eq:ulux1}) into 
(\ref{product}), and multiplying it by $D(\phi)^*$ and $\Gamma_M^*$ 
from the left and the right sides respectively, we obtain 
\begin{equation}
U_{PMNS}^{std} =   D(\gamma)~ U_L^{std} (\delta_q)~ D(\alpha) ~
U_X^{std} (\delta_X) D(\eta)\,.
\label{eq:working2}
\end{equation}
The phases  
\be
\gamma \equiv \psi - \phi, ~~~~ \eta \equiv z - \beta  
\nonumber
\ee
should be selected in such  a way that 
they bring  the RHS of \,(\ref{eq:working2}) to the standard parametrization form. 

The conditions, that the  matrix on the RH side of (\ref{eq:working2}) is  
in the standard parametrization, are given by the following 
5 equations  
\be
{\rm Arg\,} \{U_{e1} \} = 
{\rm Arg\,}\{ U_{e2}\} = {\rm Arg\,} \{U_{\mu 3}\} = {\rm Arg\,} \{U_{\tau 3}\} = 0, 
\label{eq:st-cond}
\ee 
\be
|U_{e1}|~ {\rm Im}\,U_{\mu2} = |U_{e2}|~ {\rm Im}\,U_{\mu1} \,.
\label{eq:st-cond2}
\ee
These conditions which we call the Standard Parametrization (SP) conditions 
fix 5 phases $\gamma_e, ~ \gamma_\mu, \gamma_\tau$  and $\eta_1,~ \eta_2$. 
Notice that conditions (\ref{eq:st-cond}) determine the phases of the mixing matrix up to a rephasing: 
$U_{e3} \rightarrow e^{i \Phi} U_{e3}$, and 
$(U_{\mu1},  U_{\mu2}, U_{\tau 1},  U_{\tau2}) 
\rightarrow e^{- i \Phi} (U_{\mu1},  U_{\mu2}, U_{\tau 1},  U_{\tau2})$. 
This allows, in particular,  to eliminate the phase of the 1-3 element. 
It is the condition (\ref{eq:st-cond2}) that fixes $\Phi$ and removes the ambiguity.

Once the SP-conditions are satisfied  the phase of the 1-3 element 
of the obtained matrix will give 
\be
\delta_{CP} = -{\rm Arg\,}\{U_{e3}\},~~ {\rm or}~~ \sin \delta_{CP} = -\frac{{\rm Im}\,U_{e3}}{|U_{e3}|},   
\nonumber
\ee
and the Majorana phases equal
\be
\beta = z - \eta. 
\nonumber
\ee 

\subsection{Quark-lepton similarity and general expression for the Dirac CP phase}
\label{sec:QLCgen}

The main assumption in this paper is that the Dirac mass matrix of neutrinos 
has similar structure to that of quarks: $m_D \sim m_u$ or $\sim m_d$, as can  be motivated  
by the  Grand unification or common flavor symmetry
with the same charge assignments. Consequently,   
the mixing in leptonic sector which follows from the 
Dirac matrices is similar to that in the quark sector: 
\be
U_{L}^{std}(\delta_L)\sim V_{CKM}^{\dagger}(\delta_q)\,. 
\label{eq:qlsym}
\ee
Essentially, we will only assume that mixing matrix $U_{L}$ has a hierarchical structure of 
elements, as the CKM matrix, { i.e.}, 
$V_{ud}\gg V_{cd}\gg V_{td}$, {etc.}, and express the smallness of these quantities 
by referring to the Wolfenstein parameter $\lambda$. We make no use on any other details of 
this similarity. In particular, the parameter $\lambda$ 
does not have to be exactly the same as in the quark sector.
 
According to (\ref{eq:qlsym}), we will suggestively denote the elements  
$(U_{L}^{std})_{li}$ by the elements of $V_{ul}^*$, where the charged lepton index $l=(e,\mu,\tau)$  
here corresponds to the down quarks $(d,s,b)$ in $V_{CKM}$ and the neutrino index $i=(1,2,3)$ 
corresponds to up-quarks $(u,c,t)$.
Denoting the elements of the matrix $U_{X}^{std}$ by $X_{li}$ 
we obtain for the matrix elements on the RHS of (\ref{eq:working2})
\be
U_{lj}=e^{i(\gamma_l+\eta_j)}\left[ 
V_{ul}^* X_{ej} e^{i\alpha_e} + 
V_{cl}^* X_{\mu  j} e^{i\alpha_\mu} + 
V_{tl}^* X_{\tau j} e^{i\alpha_\tau} \right]\,,
\label{eq:uli}
\ee
where $l = e, \mu, \tau$ and $j = 1, 2, 3$. We remind that in $V$ we replace $e \rightarrow d$, $\mu \rightarrow s$, 
$\tau \rightarrow b$.
 
Introducing $\xi_{lj}$ - the phases of the expressions in the brackets of (\ref{eq:uli}), 
we can rewrite the elements of the PMNS matrix (\ref{eq:uli}) as 
\be
U_{lj}=e^{i(\gamma_l+\eta_j + \xi_{lj})} |U_{lj}|. 
\nonumber
\ee
The phases $\gamma_l$ and $\eta_j$ should be  determined from  
the conditions of the standard parametrization. 

The elements $V_{ud}$, $V_{us}$, $V_{cb}$, and $V_{tb}$ are real. The elements 
$V_{cd}=-|V_{cd}|e^{i\phi_{cd}}$ and $V_{ts}=-|V_{ts}|e^{i\phi_{ts}}$ 
have an overall negative sign, 
so that the phases $\phi_{cd}$ and $\phi_{ts}$ 
are of order $\lambda^4$ and $\lambda^2$, respectively. 
The other phases are defined as usual,  
$V_{ub}=|V_{ub}|e^{i\phi_{ub}}$, 
$V_{td} =|V_{td}|e^{i\phi_{td}}$, and $V_{cs}=|V_{cs}|e^{i\phi_{cs}}$. 
The phases $\phi_{ub}$ 
and $\phi_{td}$ are ${\cal O}(1)$, while $\phi_{cs}$ is of order $\lambda^6$ and can be neglected. 
All these phases are known in terms of the quark CP violating phase $\delta_q$.

Consider the element $U_{e3}$ which contains the Dirac CP phase:  
\begin{equation}
U_{e3} =  s_{13} e^{- i\delta_{CP}} =  
e^{i\gamma_e} \left[
e^{i\alpha_e} V_{ud} X_{e3} - 
e^{i(\alpha_\mu - \phi_{cd})} |V_{cd}| X_{\mu 3} + 
e^{i(\alpha_\tau - \phi_{td})} |V_{td}| X_{\tau 3}\,
\right]. 
\label{eq:Ue3}
\end{equation}
Modulus and argument of $U_{e3}$ determine $\theta_{13}$ and $\delta_{CP}$, respectively. 
Since $|U_{e3}| = s_{13}$, from (\ref{eq:Ue3}) we obtain 
\bea
\sin \delta_{CP} & = & - \frac{1}{s_{13}}
\left[
\sin(\alpha_e + \gamma_e  -\delta_X) V_{ud} |X_{e3}| 
- \sin(\alpha_\mu + \gamma_e  - \phi_{cd}) |V_{cd}| X_{\mu 3}  
\right.
\nonumber\\
& + &  \left. \sin (\alpha_\tau + \gamma_e - \phi_{td})|V_{td}| X_{\tau 3}
\right]\,. 
\label{eq:cpphase}
\eea
Recall that the phases $\alpha_l$  and 
$\delta_X$ parametrize the CP violation which originates 
from the RH sector.  The phase $\gamma_e$ is fixed by 
the standard parametrization conditions: 
$\gamma_e = \gamma_e (\alpha_l, \delta_X, \delta_q)$. The phase $\eta_3=0$, as 
$z_3$ and $\beta_3$ can be chosen to be zero, and the above expressions 
do not explicitly depend on $\eta$.  The important feature of the result 
(\ref{eq:cpphase}) is that contribution of $\alpha_{\tau}$  to $\delta_{CP}$ 
is always suppressed by $V_{td}/s_{13} \sim \lambda^2$, 
$\delta_X$ is  suppressed by $X_{e3}$,  
whereas the contributions
of $\alpha_{e}$ and $\alpha_{\mu}$ are unsuppressed.

\section{A CKM-type origin of the leptonic CP violation}
\label{sec:Dirac}

Suppose that the only source of CP violation is 
$U^{std}_L(\delta_L)$ $\approx V_{CKM}(\delta_q)$, { i.e.}, 
the matrix of transformation of the LH  neutrino components that diagonalizes
$m_D$. This is a direct analogy to the Kobayashi-Maskawa mechanism in the quark sector,  
as previously considered e.g. in \cite{xing1}. It  corresponds to $U_X$ being a real matrix,  so that  
\be
\alpha_l=0, ~~~ z = \delta_X = 0.  
\nonumber
\ee
The matrix in front of $U^{std}_L(\delta_L)$ can be absorbed into the  
phases of the charged leptons. This can be thought of as the minimal CP violation 
that we expect for  leptons if their Dirac masses are similar to quarks. 
In the context of  the seesaw mechanism 
such a  situation 
can be realized  if both $U_R$ and $M_R$ are real, and the 
diagonal phase matrices vanish or cancel with each other. 
The cancellation  can be due certain  symmetries for RH neutrino components.  
In this case according to 
(\ref{eq:Ue3}) and (\ref{eq:cpphase}): 
\begin{equation}
U_{e3} =  
e^{i\gamma_e} \left[V_{ud} X_{e3} -
|V_{cd}| X_{\mu 3} +
e^{- i \phi_{td}} |V_{td}| X_{\tau 3}\,
\right]\,,
\label{eq:Ue3-0}
\end{equation}
and 
\begin{equation}
\sin \delta_{CP} = - \frac{1}{s_{13}} \left[
\sin \gamma_e (V_{ud} X_{e3} - |V_{cd}| X_{\mu 3})  +
   \sin{(\gamma_e - \phi_{td})}|V_{td}| X_{\tau 3} \right]+{\cal O}(\lambda^4),
\label{eq:cpphase1}
\end{equation}
where $\phi_{cd}$ has been neglected.
The  absolute value   of $U_{e3}$  according to 
(\ref{eq:Ue3-0}) equals 
\be
|U_{e3}| = s_{13} = |A| \equiv \big|V_{ud} X_{e3} - |V_{cd}| X_{\mu 3}\big| + {\cal O}(\lambda^3). 
\nonumber
\label{comb}
\ee
Therefore 
\begin{equation}
\sin \delta_{CP} = - {\rm sign} \{ A \}  \sin \gamma_e  - 
   \frac{1}{s_{13}} \sin{(\gamma_e - \phi_{td})}|V_{td}| X_{\tau 3}.
\nonumber
\end{equation}
Thus,  the CP phase is determined essentially by $\gamma_e$ which 
we find (see Appendix A for details) by imposing  the SP conditions 
(\ref{eq:st-cond}, \ref{eq:st-cond2}) to be  
\be
\gamma_e =  \frac{X_{e1}^2X_{\mu 2}X_{\tau 2}
- X_{e2}^2X_{\mu 1}X_{\tau 1}}{V_{ud} X_{e1} X_{e2} X_{\tau 3}} s_{13}^q \sin \delta_q 
+ {\cal O}(\lambda^4)\, , 
\label{eq:ge1}
\ee
where we used the result (\ref{eq:ge}) and 
$|V_{td}| \sin \phi_{td} \equiv {\rm Im}\,V_{td} = s_{13}^q \sin \delta_q$.  
Since $s_{13}^q = \lambda^3$ the expression 
(\ref{eq:ge1}) shows that $\sin\gamma_e = {\cal O}(\lambda^3)$.

Let us express the elements $X_{li}$ in terms of the elements 
of $U_{PMNS}^{std}$. Using the relations (\ref{eq:uli}), 
at ${\cal O}(1)$  we have $X_{lj} \approx |U_{lj}|/ (V_{CKM})_{ll}$, while  
$X_{e3}$ turns out to be of the order $\lambda$:  
$X_{e3} = \pm s_{13}/V_{ud} + s_{23}|V_{cd}|/(V_{ud}|V_{cs}|)$.
With these expressions for $X_{li}$ and $\gamma_e$, we obtain from (\ref{eq:cpphase1})
\be
\sin \delta_{CP} = - \sin\delta_q \frac{s_{13}^q}{s_{13}} c_{23}  
  \left[1 + 2 s_{13} \tan \theta_{23} \cot 2 \theta_{12}\right] + 
{\cal O}(\lambda^4, \lambda^3 s_{13})\, . 
\label{dddd}
\ee
Similarly according to  (\ref{eq:beta12a}) and (\ref{eq:beta12b}), the Majorana phases are
\bea
\beta_1&=&    
\frac{s_{23}c_{12}}{s_{12}}s_{13}^q \sin\delta_q + {\cal O}(\lambda^4)\,,\nonumber\\
\beta_2&=& - \frac{s_{23}s_{12}}{c_{12}}s_{13}^q \sin\delta_q 
+ {\cal O}(\lambda^4)\,.\nonumber
\eea
The following comments are in order. 

1. The main term in (\ref{dddd}) is of the order $\lambda^3/ s_{13} \sim \lambda^2$,  
that is,  suppressed by $\sim \lambda^2$.  
This agrees with results obtained previously (e.g., \cite{xing1}, \cite{yasaman}). 
At leading order (\ref{dddd}) can be rewritten as 
\be
s_{13} \sin \delta_{CP} = (- c_{23}) s_{13}^q \sin \delta_q,  
\nonumber
\ee
or $ {\rm Im}\,  U_{e3}  = - c_{23} {\rm Im}\,V_{ub}$. 
So, the Dirac CP phase in the leptonic sector is suppressed 
because the mixing is relatively large, compared to quark mixing. 

2. The subleading term in the Dirac CP phase is of the order 
$\lambda^3$, and it is proportional to deviation of the 
2-3 mixing from maximal.  

3. Numerically we have 
$\sin \delta_{CP} \approx - 0.05 \sin\delta_q = -0.046$, as $\delta_q=1.2\pm0.08$ radian. 
To determine the phase itself we should also estimate $\cos \delta_{CP}$.  
Since $\sin \delta_{CP} \ll 1$, we have
$\cos \delta \approx \pm 1$.
Therefore according to (\ref{eq:Ue3-0})
$\cos \delta_{CP} =  {\rm sign} \{A\}$, 
which corresponds to either
\be
\delta_{CP} \approx -\delta~~{\rm or}~~\delta_{CP}\approx\pi+\delta\,,
\nonumber
\ee 
where the deviation $\delta \approx  (s_{13}^q /s_{13}) c_{23} \sin\delta_q$, 
is of the order $\lambda^2$. 

4. The Majorana phases are smaller and suppressed as $\lambda^3$. 
Numerically one finds that $\beta_1\approx0.01$ and $\beta_2\approx-0.005$.
Notice that these  are the ``induced" phases  
by the Dirac quark phase $\delta_q$ and SP conditions.
Indeed, the  phase  $\delta_q$ appears in a mixing matrix that is not in the standard form,   
and $\beta_i$ are the phases obtained in rephasing procedure to bring this matrix to the standard form.

5. As we remarked before, the Dirac phase can be obtained from  
the Jarlskog invariant in the standard parametrization: 
\be
J_{CP} \equiv {\rm Im} \left[U_{e1}^* U_{\mu 3}^* U_{e3} U_{\mu 1} \right] = \frac{1}{8} 
\sin 2 \theta_{13} \sin 2 \theta_{13} \sin 2 \theta_{13} \cos \theta_{13} \sin \delta_{CP}.  
\label{yarl}
\ee
Using expressions (\ref{eq:uli}) for the elements in the LHS of this equality 
taken for all zero phases but $\delta_q$ we obtain in the lowest order 
\be
J_{CP} = - V_{cs}^2 V_{ud} X_{\mu 1} X_{\mu 2} X_{\mu 3}~ {\rm Im} V_{td}
\approx - X_{\mu 1} X_{\mu 2} X_{\mu 3}~ {\rm Im} V_{td}. 
\nonumber
\label{jar2}
\ee
Expressions  (\ref{eq:xspl4}) and (\ref{eq:xspl9}) in Appendix B, 
for $X_{\mu i}$ in terms of PMNS mixing angles  
allow to rewrite this as
\be
J_{CP} = s_{12} c_{12} s_{23} c_{23}^2  {\rm Im} V_{td}
=  s_{12} c_{12} s_{23} c_{23}^2  s_{13}^q \sin \delta_q . 
\nonumber
\ee
Finally, inserting this into LHS of  eq. (\ref{yarl}) we find 
$\sin \delta_{CP} = - c_{23} (s_{13}^q /s_{13}) \sin \delta_q$ 
which coincides with the lowest order term in eq. (\ref{dddd}).

6. The results obtained in this section do not actually depend on 
mechanism of neutrino mass generation. They are based on a general parametrization of the 
PMNS matrix (\ref{product}), with the assumption that $U_{L} \sim V_{CKM}^{\dagger}$ is 
the only source of the CP violation and requirement that the product (\ref{product}) 
reproduces the observed lepton mixing angles. 
Although we have motivated this ansatz in the context of seesaw type I, 
any model that satisfies $U_{L} \sim V_{CKM}^{\dagger}$ and has 
no other source of CP violation leads to the same result.

\section{General Case of CP Violation}
\label{sec:general}

In general the assumption made in the previous section, that the left transformation is the only 
source of CP violation, is not valid for Majorana neutrinos implied by seesaw.  
In the case of Majorana neutrinos, phases of 
the RH sector become important for PMNS mixing. 
In particular, the CP phase in the right
matrix $U_{R}$ will contribute to $\delta_{CP}$. 
The CP violation in RH sector doesn't affect the CP violation in the CKM matrix  
because quarks do not have a Majorana mass term. In this sense, the analogy between 
the lepton and quark sector cannot be exact even if Dirac matrices are the same - 
the matrix $U_R$ has physical consequences for neutrinos.

Consider the most general possibility, when  
CP violating parameters exist in both the Dirac 
and Majorana mass matrices involved in the seesaw. 
Neglecting terms of the order $\lambda^3$ we obtain from 
(\ref{eq:cpphase})
\be
\sin \delta_{CP} =
-  \frac{1}{s_{13}}
\left[
\sin(\alpha_e + \gamma_e -\delta_X) V_{ud} |X_{e3}| - 
\sin(\alpha_\mu + \gamma_e) |V_{cd}| X_{\mu 3}\right]\,. 
\label{gensol}
\ee
In the leading order in $\lambda$ 
the conditions of standard parametrization 
(\ref{eq:st-cond}) give 
\be
\eta_{1} + \alpha_e + \gamma_e = 0, ~~~~
\eta_{2} + \alpha_e + \gamma_e = 0\,,
\label{etasol}
\ee
\be
\gamma_{\mu}=-\alpha_{\mu}, ~~~~
\gamma_{\tau}=-\alpha_{\tau}, 
\nonumber
\ee
and the  5th condition reads: 
\be
\sin (\gamma_\mu + \eta_2) = r \sin (\gamma_\mu + \eta_1). 
\label{fifth}
\ee
(Notice that $r \approx - 2$, because $U_X$ is close to being $U_{TBM}$). 
The only solution of this system is the following: 
$\eta_{1} = \eta_{2} \equiv \eta$ from (\ref{etasol}), then 
$\gamma_\mu = -  \eta$ from (\ref{fifth}), and then 
$\alpha_\mu = \eta$,  $\gamma_e + \alpha_e = - \eta$. 
Inserting these expressions into (\ref{gensol})
we obtain 
\be
\sin \delta_{CP} =
  \frac{1}{s_{13}}
\left[ V_{ud} |X_{e3}| \sin(\alpha_\mu + \delta_X)  -
|V_{cd}| X_{\mu 3} \sin \alpha_e   \right]\,.
\nonumber
\label{gensol1}
\ee
All three phases $\delta_X$, $\alpha_e$, and $\alpha_\mu$ 
are free parameters and one can obtain any value of the CP phase. 
In specific cases, some of these phases can be 
removed or fixed resulting in a more precise prediction, e.g., if  $X_{e3} = 0$,   
we get $\sin\delta_{CP}\approx -\sin\alpha_e$.
For $\alpha_e = \alpha_\mu =  \delta_X = 0$ we obtain 
$\delta_{CP} = 0$, in agreement with our consideration in Sec.\,\ref{sec:Dirac} at this order.

If $\alpha_\mu \neq 0$ 
and $\alpha_e =  \alpha_\tau = 0$,  we obtain by using 
the standard parametrization conditions 
\be
\sin \delta_{CP} = \sin \alpha_\mu 
\frac{V_{cd} X_{\mu 3}}
{|- V_{cd} X_{\mu 3}e^{i \alpha_\mu} + V_{ud} X_{e3}|}. 
\nonumber
\ee
According to this expression  $\delta_{CP}$ can be of the order 1 if $\alpha_\mu$ is unsuppressed.

The Majorana phases equal  
\be
\beta_1 = z_1 - \alpha_\mu, ~~~ \beta_2 = z_2 - \alpha_\mu,  
\nonumber
\ee
which gives  $\beta_1 - \beta_2 = z_1 -  z_2$, 
where $z_i$ are also unknown parameters, which can be fixed once $M_R$ is determined. So, in general, 
all leptonic CP phases are unconstrained and can be large.\\

\section{CP violation from $U_R$ and Seesaw Enhancement of the CP phase}
\label{sec:UR}

\subsection{CP phases in the Left-Right symmetric case}

Here we explore the minimal extension 
of the CKM case  that includes effect of the RH sector. 
In the spirit of L-R symmetric models we assume that 
\be 
{U}_{R} \approx U_L \sim V_{CKM}^{\dagger}, 
\nonumber
\ee 
and there is no CP violation in $M_R$ in the L-R symmetry basis.
So, 
\begin{equation}
M_X \equiv - m^{diag}_{D} V_{CKM} M_{R}^{-1} V_{CKM}^T  m^{diag}_{D},
\label{eq:mx3}
\end{equation}
where now $M_X$ is a complex symmetric matrix. 
The CP violation in $U_R \sim  U_L$ is very small, being suppressed by $\lambda^3$.

To elucidate the role of CP violation from  $U_R$ and effect of seesaw 
we assume  that $M_R$ has the following form: 
\begin{equation}
M_{R}^{-1} = V_{CKM}^{0 T}  (m^{diag}_{D})^{-1} \tilde{M}_{TBM} 
(m^{diag}_{D})^{-1}  V_{CKM}^{0}\,, 
\label{eq:mr-anz}
\end{equation}
where $V_{CKM}^{0} = V_{CKM} (\delta_q = 0)$ is 
the CKM-like matrix with zero value of the CP phase and $\tilde{M}_{TBM} \approx {M}_{TBM}$. 
The latter ensures that matrix $U_X$ is close to $U_{TBM}$, 
which leads to the observed PMNS mixing angles.

Inserting expression (\ref{eq:mr-anz})  into (\ref{eq:mx3}) we can represent $M_X$ as 
\be
M_X = - K \tilde{M}_{TBM} K^T,  
\nonumber
\ee
where 
\be
K \equiv m^{diag}_{D} V_{CKM} V_{CKM}^{0 T}  (m^{diag}_{D})^{-1}
\nonumber
\ee
is the correction matrix that captures the effect of a non-zero CP phase.   
Indeed, for $\delta_q = 0$, $K = I$ the above would provide 
$M_X\approx-\tilde{M}_{TBM}$.  

Computing explicitly, we find
\be
V_{CKM} V_{CKM}^{0 T} = I + 
\left(
\begin{array}{ccc}
0 & 0 & - V_{td}^*\\
0 & 0 &  0\\
V_{td} & 0 &  0
\end{array}
\right), 
\nonumber
\ee
and $V_{td} \approx \lambda^3 (1 - e^{i\delta}) \equiv \lambda^3 \xi$. 
Let us take $m_D = m_{3D}\, {\rm diag} (\lambda^m , \lambda^n, 1)$. 
We can also include coefficients of order one here, but they will not change final 
conclusion.
Then 
\be
K =    
\left(
\begin{array}{ccc}
1 & ~0~ & - V_{td}^* \lambda^m\\
0 & 1 &  0\\   
V_{td}\lambda^{-m} & 0 &  1
\end{array}
\right) \approx 
\left(
\begin{array}{ccc}
1 & 0 & 0 \\
0 & ~~1~~ &  ~0~\\
\xi \lambda^{3 - m} & 0 &  1
\end{array}
\right).  
\label{matK}
\ee
For $m \geq 4$, the  3-1 element of the correction matrix 
is large, i.e., enhanced as $ \geq \xi \lambda^{- 1}$.  
It is this factor, related to the strong hierarchy of the 
eigenvalues of the Dirac matrix, that can lead to 
enhancement of the CP violation. Note that the correction 
in (\ref{matK}) does not depend on the second eigenvalue $\lambda^n$. 

We take 
\be
\tilde{M}_{TBM} \sim 
m_0
\left(
\begin{array}{ccc}
a \lambda^p & b \lambda &  f \lambda\\
... & 1 &  g\\
... & ... &  h
\end{array}
\right),  
\label{eq:tbm}
\ee
where $a, b, g, h$ are real coefficients of the order 1. 
Then using the correction matrix (\ref{matK}) we obtain   
\be
M_X \propto    
\left(   
\begin{array}{ccc}
~a \lambda^p~ & b\lambda &     a \xi   \lambda^{-m + p +3}  + f \lambda \\
...  & 1 & b \xi \lambda^{-m + 4}  + g  \\
... & ... & ~~a \xi^2 \lambda^{- 2m + p + 6} + 2 f \xi \lambda^{-m + 4 } + h~~
\end{array}
\right)\,.    
\label{eq:mxx}
\nonumber
\ee 
The only possibility to have $M_X$ be an approximate TBM mass matrix is 
$m \leq 4$ and  $p \geq 2$.  That is, the 
hierarchy of the Dirac mass matrix is strongly restricted by the condition that 
correct PMNS mixing is reproduced.  
If the hierarchy 
of the eigenstates of the Dirac mass matrix is too strong, {i.e.}, $m >  4$, no solution 
which gives correct mixing angles exists in the presence of a CP violating phase. 
At the same time a solution always exists for arbitrarily strong hierarchies if there is 
no CP phases in $U_R$.

Taking  $m=4$ and $p=2$, we obtain  
\be
M_X \propto
\left(
\begin{array}{ccc}
a \lambda^2 & ~~b\lambda~~ &  \lambda (a \xi + f) \\
...  & 1 &  b \xi + g   \\
... & ... & ~~a \xi^2  + 2 f \xi +  h  
\end{array}
\right),  
\label{eq:mxx1}
\ee
and  corrections to all the elements are suppressed by at least $\lambda^2$. 
Now the problem is to find phases of the matrix $U_X$ ($y_l, \delta_X, z_i$) 
that diagonalizes (\ref{eq:mxx1}). 

\subsection{Factorization of phases}
\label{sec:factorization}

The phases of $U_X$  can be found  immediately 
if the phases  are factorized from $M_X$.  Under the conditions 
\be
{\rm Arg}\,(a \xi + f) = {\rm Arg}\,(b \xi + g) = 
\frac{1}{2}{\rm Arg}\,(a \xi^2 + 2f\xi + h) \equiv \phi_F \, , 
\nonumber
\ee
which we will call the phases factorizations conditions, the matrix 
(\ref{eq:mxx1}) can be written as 
\be
M_X =  D(\phi_F) M_X^0  D(\phi_F), 
\nonumber
\ee
where $D(\phi_F) = (1, 1, e^{i\phi_F})$ and  
\be
M_X^0 \propto
\left(
\begin{array}{ccc}
a \lambda^2 & b \lambda &   \lambda a |F|  \\
...  & 1 &  b |F|   \\
... & ... & a |F|^2
\end{array}
\right)\,.
\label{eq:mxx3}
\ee
Here $F \equiv |F| e^{i \phi_F}  \equiv   \xi  + f/a$. 

The factorization conditions can be satisfied if 
\be
\frac{f}{a} = \frac{g}{b}, ~~~~ f^2 = a h.
\nonumber
\ee

Since $M_X^0$ is real and, in general,  can be diagonalized by real matrix $O$,  
we have $U_X = D(\phi_F) O$. So that in the notation of (\ref{eq:ulux}), 
$y_e = y_\mu = \delta_X = z = 0$ and $y_\tau = \phi_F$. Furthermore,  since in the L-R 
symmetric case $D(\xi)$ is irrelevant, $\alpha_l = y_l$ and $\alpha_{\tau} = y_\tau = \phi_F$. 
Thus,  $D(\alpha) = {\rm diag} (1, 1, e^{i \alpha_\tau})$ 
and  the factorization phase  is determined by 
\be
\tan \alpha_\tau = \frac{\sin \delta_q}{1  +  f/a - \cos \delta_q}.  
\nonumber
\label{alphatau}
\ee
Furthermore,  
\be
|F| = |\xi +  f/a| = \left[ (1 + f/a)^2 - 
2(1 + f/a) \cos\delta_q  + 1 \right]^2 . 
\nonumber
\ee
If $f  = -1$ and  $a   = 1$ 
we have $\alpha_\tau  = - \delta_q$ and $|F| = 1$. 
More generally, for the interval $a = 1 - \lambda$ to $1 + \lambda$ we obtain   
that $\alpha_\tau$  changes from  $-74^{\circ}$ to $- 86^{\circ}$.  
For $f > 0$ the interval for the  phase $\alpha_\tau$ is  $27^{\circ} -  30^{\circ}$. 
In both cases $\alpha_\tau$  differs from $\phi_{td} \approx 50^{\circ}$. 
Solving the SP conditions, (see Appendix A) 
we find 
\be
\sin \delta_{CP} \approx - {\rm sign}\{A\} 2 \sin \phi_{td} V_{ts}  
\cot  2 \theta_{23}  - \frac{1}{s_{13}} 
\sin (\alpha_\tau - \phi_{td}) |V_{td}| X_{\tau 3}. 
\label{cptau}
\ee
Thus, in the case of factorization with only $\alpha_\tau \neq 0$, the final value of the CP  
phase is still small, being  suppressed by $\lambda^2$.  The reason is that 
$\alpha_\tau$ enters the expression for $\sin \delta_{CP}$ with small factor 
 $V_{td}$. 
The Majorana phases (which appear as by-product of the standard parametrization 
conditions) equal (see Appendix B)
\be
\sin \beta_1 \approx \sin \beta_2 \approx - {\rm sign}\{A\} 2 \sin \phi_{td} V_{ts} \cot  2 \theta_{23}.  
\nonumber
\ee
For $\alpha_\tau = \phi_{td}$ all three CP phases are equal.

The matrix  (\ref{eq:mxx3}) does not satisfy the exact TBM conditions:    
\be
(M_X)_{12} = -(M_X)_{13}, ~~~~ (M_X)_{13} = (M_X)_{33}, ~~~~ 
(M_X)_{22} - (M_X)_{23} = (M_X)_{11} + (M_X)_{12},   
\nonumber
\ee
which for (\ref{eq:mxx3}) take the form 
\be
b = - a |F|, ~~~ a |F|^2 = 1, ~~~~ 1 - b |F| \approx b \lambda.  
\nonumber
\ee
Indeed, from the first and second equalities we have  $b |F| = - 1$ and from the last one:  
$ b |F| \approx  1 - b \lambda \approx 1$. The deviation of $M_X^0$ from the TBM form  leads, 
in particular, to a non-zero 1 - 3 mixing: 
\be 
X_{e3} \approx \tan \theta_{13}^X  \sim \frac{\lambda}{ \sqrt{2}} \frac{1}{|F|}
\nonumber
\ee    
which can be in agreement with data.

\subsection{Seesaw enhancement of CP violation}
\label{sec:enhanced}

In general in the absence of factorization 
the mass matrix $M_X$ will generate a non-zero $\alpha_e$, $\alpha_\mu$, and $\delta_X$,
and consequently  a large $\delta_{CP}$. 
Expressions for phases of $U_X$ in the three generation case are 
very complicated and difficult to analyze. Therefore to show effect of  
enhancement of the CP phases we will consider the two leptonic generations. 
In the case of a hierarchical neutrino mass spectrum  the  
$2 - 3$ block of elements in the mass matrix is dominant, 
with elements of the first row and column being suppressed by 
$m_2/m_3 \sim  \lambda$ as in (\ref{eq:mxx1}). Therefore we consider the second 
and third neutrinos. Results obtained in this approximation are expected to receive corrections 
of the order $\lambda$ when  mixing with the first neutrino is turned on. 

The matrix $M_X$ can be written as 
\be
M_X = D(\Phi_H) M_X^0 D(\Phi_H), 
\nonumber
\ee
where  $D(\Phi_H) = {\rm diag} (1, e^{i\Phi_H/2})$ and 
\be 
M_X^0 = m_0
\left(
\begin{array}{cc}
1   &  G e^{i\psi}   \\
... & H
\end{array}
\right).   
\label{eq:m2d}
\ee 
Here $G e^{i\Phi_G} \equiv  b \xi + g $, 
$H e^{i\Phi_H} \equiv a( \xi^2  + 2 f \xi +  h)$,  and 
$\psi \equiv \Phi_G - \Phi_H/2$. It is easy to show that selecting parameters 
$a, b,  g,  f, h$ one can get any value of $\psi$ from zero to 
${\cal O}(1)$. 

We will diagonalize $M_X^0$ (\ref{eq:m2d}) with 
$U^0_X = D(y^0) {R}_X(\theta) D(z)$,  where  ${R_X}(\theta)$  is a
$2 \times 2$ rotation matrix, 
$D(y) = {\rm diag} (e^{i y_\mu}, e^{i y^0_\tau})$, and 
$D(z) = {\rm diag} (e^{i \beta_2}, 1)$ are the phase matrices.
Then $U_X = D(\Phi_H) U_X^{0}$.  
The diagonalization condition   
$U_X^{0 \dagger} M_X^0  U_X^{0 *} = m_\nu^{diag}$, can be written 
as 
\be 
R^T_X(\theta) D(\Delta) M_X^0 D(\Delta) R_X(\theta) = e^{2 i y_\mu} D(z) m_\nu^{diag} D(z),   
\label{diagcond}
\ee
where $D(\Delta) \equiv {\rm diag}(1 , e^{i\Delta})$ and 
$\Delta \equiv y_\mu - y_\tau^{0}$. 
From (\ref{diagcond})  we obtain the relations which determine the phases 
$y_\mu,~  y_\tau^{0},~ \beta_2$:

\bea 
 \frac{1}{2}\sin 2\theta \left(1 -  H e^{i2\Delta} \right)  +  \cos 2\theta G e^{i(\psi + \Delta)} 
= 0\,, 
\nonumber\\
 c^2 + s^2 H e^{i2\Delta}  -  \sin 2\theta G e^{i(\psi + \Delta)} 
= \frac{m_2}{m_0} e^{i2(y_\mu + \beta_2)}\,,
\nonumber\\ 
 s^2 + c^2 H e^{i2\Delta}  +  \sin 2\theta G e^{i(\psi + \Delta)} 
= \frac{m_3}{m_0} e^{i2y_\mu}\,. 
\label{diageq}
\eea

The solution is very simple in the case of maximal mixing: $ \cos 2\theta = 0$, 
when the first equation in (\ref{diageq})
is satisfied for  $H = 1$  and $\Delta = 0$, so that  $y_\mu = y_\tau^{0}$. 
The two other equations give 
\be 
1  -  G e^{i\psi}  =  \frac{m_2}{m_0} e^{i2(y_\mu + \beta_2)} , ~~~~
1  +  G e^{i\psi}  =  \frac{m_3}{m_0} e^{i2y_\mu}. 
\nonumber
\ee
From these equations we obtain  
\be
\sin 2 y_\mu =   G \frac{m_3}{m_0}\sin \psi = 
\frac{G \sin \psi}{\sqrt{1 + 2 G\cos \psi +G^2}},  
\label{eqyyy}
\ee 
and $G$ determines the mass hierarchy: 
\be
\frac{m_2}{m_3} = \sqrt{\frac{1 - 2 G\cos \psi +G^2}{1 + 2 G\cos \psi +G^2}}. 
\nonumber
\ee
The equality (\ref{eqyyy}) implies that  
$\sin 2 y_\mu$ is of the order $\sin \psi$. 
And since  $\psi$ can be  
${\cal O}(1)$, can have  a large $\alpha_\mu=y_\mu$,  
and consequently,  a large $\delta_{CP}$. Furthermore, by selecting $G$  
the correct mass hierarchy can be obtained.

In the case of deviation of 2-3 mixing from  maximal,  $H \neq 1$,  one obtains in general 
corrections to the obtained results of the order $(H - 1)$. In special case  
 $\cos \psi \approx 0$ the corrections can be enhanced.

\subsection{CP phases with other assumptions on $M_R$}
\label{sec:otherass}

Similar results can be obtained with other ansatzes for $M_R^{-1}$. 

1) Consider 
\begin{equation}
M_{R}^{-1} =  (m^{diag}_{D})^{-1} \tilde{M}_{TBM}
(m^{diag}_{D})^{-1} \,,
\nonumber
\label{eq:mr-anz2}
\end{equation}
with $\tilde{M}_{TBM}$ given in  (\ref{eq:tbm}). 
It differs from the ansatz in Sec.\,\ref{sec:enhanced} by the absence of the rotation 
$V_{CKM}^{0}$.  
Taking $m_D = {\rm diag} (\lambda^4, \lambda^2, 1)$,  which  the only possibility which can lead to nearly TBM 
mass matrix for $M_X$, we obtain 
\be 
M_X \propto
\left(
\begin{array}{ccc}
a \lambda^p &  - a \lambda^p  +  b\lambda &   a \lambda^{p-1}\xi + \lambda (f - b) \\
...  & a \lambda^{p - 2} - 2b + 1 & - a \lambda^{p - 2} +  b \xi  - (f - b)  + g - 1   \\
... & ... & a \lambda^{p - 2} \xi^2  + 2 \xi (f - b)  - 2 g +  h + 1
\end{array}
\right).
\nonumber
\ee 
In contrast to the previous case, now it is possible to have $p = 1$,  
leading to dominance of terms with $a$. That is, the whole matrix 
at the lowest order is generated by the $1-1$ element of $\tilde{M}_{TBM}$:  
\be
M_X \propto
\frac{a}{\lambda}  
\left(
\begin{array}{ccc}
\lambda^2 &  - \lambda  &  \lambda \xi \\
...  & 1 & -  \xi   \\
... & ... & \xi^2 
\end{array}
\right) + 
\left(
\begin{array}{ccc}
0 &  - a b\lambda &  \lambda (f - b) \\
...  & - 2b + 1 &  b \xi - (f - b)  + g - 1   \\
... & ... &  2 \xi (f - b)  - 2 g +  h + 1
\end{array}
\right).
\nonumber
\ee
At the lowest order (the first term)  phase factorization occurs automatically and the matrix 
$M_X$ is close to TBM, having only one nonzero mass eigenvalue. 
The factorization phase equals $\alpha_\tau={\rm Arg}\,\xi = \phi_{td}$,     
and according to (\ref{cptau}) 
\be
\sin \delta_{CP} = - {\rm sign}\{A\} 2 \sin \phi_{td} V_{ts} \cot 2 \theta_{23}.
\nonumber
\label{cptau1}
\ee  

Corrections of the order $\lambda$ then generate lighter masses 
giving naturally $m_2/m_3 = O(\lambda)$ as well as  modify CP phases. 
Selecting $g$ and $h$ one can achieve phase factorization 
of the whole matrix. In this case the elements of the third column become 
\be
(M_X)_{e\tau} = a \xi^{\prime}, ~~~
(M_X)_{\mu \tau} = - \left(\frac{a}{\lambda} - b \right) \xi^{\prime}, ~~~
(M_X)_{\tau \tau} = \frac{a}{\lambda} \xi^{\prime},
\nonumber
\ee
with 
\be
\xi^{\prime} = \xi + \lambda \frac{f - b}{a}. 
\nonumber
\ee
The latter gives $\alpha_\tau = \phi_{td} + O(\lambda)$.\\

2) Instead of $U_R = V_{CKM}^{\dagger}$ we could use a more general expression 
$U_R =  D^*(\kappa) V_{CKM}^{\dagger} D(\kappa)$, where  
$D(\kappa) = {\rm diag} (e^{i\kappa_1}, e^{i\kappa_2},  e^{i\kappa_3})$.   
We can  fix $\kappa_i$ in such a way that the 3-1 element 
in the matrix $V_{CKM} D(\kappa) V_{CKM}^{0T}$, which led to the seesaw enhancement, 
is zero. For  $\kappa_1 = \kappa_2 = 0$   and $\kappa_3 = \delta_q$ we obtain
\be
V_{CKM} D(\kappa) V_{CKM}^{0 T} = 
\left(
\begin{array}{ccc}
1 & 0 & 0 \\
0 & 1 &  - \lambda^2 \xi  \\
0 &  - \lambda^2 \xi   &  e^{i\delta_q}
\end{array}
\right).
\nonumber
\ee
Through this rephasing  we moved the CP phase from the 1-3  to the 2-3 element. 
For the correction matrix we find  
\be
K = 
\left(
\begin{array}{ccc}
1 & 0 & 0 \\
0 & 1 &  - \lambda^{n + 2} \xi  \\
0 &  - \lambda^{- n + 2} \xi   &  e^{i\delta_q}
\end{array}
\right).
\nonumber
\label{matK2}
\ee
Notice that now  the second eigenvalue of $m_D$  matters.  Finally,  
with  $\tilde{M}_{TBM}$  from 
(\ref{eq:tbm}) we obtain   

\be
M_X = K  \tilde{M}_{TBM} K^{-1} 
\propto
\left(
\begin{array}{ccc}
a \lambda^p & b \lambda &   - b \xi \lambda^{-n  + 3} + f \lambda e^{i\delta_q} \\
...  & 1 & - \xi \lambda^{- n + 2}  + g e^{i\delta_q}  \\
... & ... & ~~~\xi^2 \lambda^{- 2n + 4} - 2 g  e^{i\delta_q} \xi \lambda^{-n  + 2 } +  h e^{i2\delta_q}
\end{array}
\right). 
\nonumber
\label{eq:mxxx}
\ee
$M_X \sim M_{TBM}$ can be obtained for $n = 2$. In this case   
\be
M_X  
\propto
\left(
\begin{array}{ccc}
a\lambda^p & b\lambda &   - b\lambda \xi + f\lambda e^{i\delta_q}  \\
...  & 1 & - \xi + g e^{i\delta_q}  \\
... & ... & ~~~\xi^2 - 2 g e^{i\delta} \xi  + h e^{2i\delta_q} 
\end{array}
\right).
\nonumber
\label{eq:mxxx1}
\ee
The factorization is absent, in general, but it can be achieved  by imposing relations  
$g^2 = h$, $f/b = g$. As a result,  
\be
M_X
\propto
\left(
\begin{array}{ccc}
a\lambda^p & b\lambda &   - b\lambda \xi ''    \\
...  & 1 & - \xi '' \\
... & ... & (\xi '')^2 
\end{array}
\right)\,,
\nonumber
\ee
where $\xi '' \equiv \xi - g e^{i\delta_q}$.  
If $g = -1$, we have $\xi '' = 1$. In this case 
the contribution to the CP phase from the RH sector disappears 
and we revert to the situation described in Sec.\,\ref{sec:Dirac} with CKM  
origin of CP violation.

Three main results emerge from this analysis of CP violation 
under the assumptions that $U_L\approx U_R\sim V_{CKM}^\dagger$ and there is 
no CP violation in $M_R$ in the L-R symmetric basis: 

1. The hierarchy of Dirac masses of neutrinos cannot be too strong, 
{i.e.}, $m_{1D}/m_{3D}\leq \lambda^4$ and $m_{2D}/m_{3D}\leq \lambda^2$. 
The observed mixing angles of $U_{PMNS}$ impose this requirement. 
This is significantly weaker than the mass hierarchy of up quarks.

2. The CP phases can in general be large, 
even if the only sources of CP violation are the Dirac phases in $U_L^{std}$ 
and $U_R^{std}$, where the CP phase effect  is suppressed by $\lambda^3$. 
This enhancement originates from  seesaw and the hierarchy of Dirac masses of neutrinos.

3. If parameters of  $M_R$ satisfy  certain relations 
-- the phase factorization conditions (which could  be a consequence of  
some  symmetry), the phases can factor out from $M_X$. Furthermore,    
the only non-vanishing phase which enters the phase factors is  $\alpha_\tau$. 
This is related to certain pattern of CP violation in CKM matrix. 
In this case  no enhancement occurs and $\delta_{CP}$ turns out 
of the order $\lambda^2$. 

$M_X$ deviates from $M_{TBM}$ since the correction in $K$ is relatively large: 
being of the order $\lambda^3$,  which is still larger than the hierarchy of masses in $m_D$.

\subsection{Enhancement of a small phase in $U_R$}
\label{sec:maxenhanced}

In the previous examples large  $\delta_{CP}$ has been obtained at the cost of deviation 
of $U_{X}$ from $U_{TBM}$.  With decrease  of $\delta_q$, correction 
to the matrix $M_X$ due to CP violation (given by $K$) decreases and 
$M_X \rightarrow  \tilde{M}_{TBM}$.  So, $M_X$ can  coincide with $M_{TBM}$ up to small 
corrections.  (This however implies that we  depart from L-R symmetry or quark-lepton 
similarity, assuming smaller values of $\delta_q$.)

Suppose $\delta_q = \epsilon \lambda^2$,  where $|\epsilon| \leq 1$. In this case 
$\xi \approx - i \delta_q = - i \epsilon \lambda^2$ and  
\be
K=
\left(
\begin{array}{ccc}
1 & 0 & 0 \\
0 & ~~1~~ &  ~0~\\
-i \epsilon \lambda & 0 &  1
\end{array}
\right)\,.
\nonumber
\ee
Here the  correction is suppressed by $\lambda^2$ in comparison with that in (\ref{matK}). 
Let us take for definiteness  the parameters 
of $\tilde{M}_{TBM}$ to be $a = b = f = - g = h = 1$ which ensures the exact 
TBM mixing in the lowest order with vanishing lowest neutrino mass. Then 
\be
M_X= 
m_0
\left(
\begin{array}{ccc}
\lambda & \lambda & \lambda - i \epsilon \lambda^2\\
... & 1+\lambda &  -1+\lambda - i \epsilon\lambda^2\\
... & ... &  1+\lambda- 2i \epsilon \lambda^2
\end{array}
\right)\,,
\nonumber
\ee
where $\lambda\approx\sqrt{\Delta m^2_{21}/\Delta m^2_{31}}$ leads to the correct 
neutrino masses. The additional imaginary terms  
give  corrections to the TBM values of  the 
$1-2$ and $2-3$ mixing angles proportional to $\epsilon \lambda^2$.
They also generate small  $1-3$ mixing: 
$X_{e3} \approx \epsilon \lambda^2$ and  
\be
\delta_X\approx \frac{\pi}{2}+{\cal O}(\epsilon \lambda^2)\,.
\nonumber
\ee
All the other induced phases are close to 0 or to $\pi$, { i.e.}, $D(y)={\rm diag}(1,-1,1)$ 
and $D(z)={\rm diag}(-1,-1,0)$, with corrections as $\epsilon \lambda^2$.
According to (\ref{eq:cpphase}) this  contributes 
to the PMNS phase as 
\be
\sin \delta_{CP} \approx - \frac{X_{e3}}{s_{13}} \approx - \frac{\epsilon \lambda^2}{s_{13}} \sim \epsilon \lambda. 
\nonumber
\ee
So, seesaw can convert a tiny CP phase $\delta_q\equiv \epsilon\lambda^2$ 
in $U_R$  to a maximal CP phase $\delta_X\approx \pi/2$ in $U_X$. 
This happens  because of the large hierarchy 
of Dirac masses and  seesaw.

\section{Remarks on Phenomenology}
\label{sec:consequences}

Our results have the following phenomenological consequences: 

\begin{enumerate}
\item For the scenarios with CKM type CP violation  and in the L-R symmetric case with 
phase factorization the value of  $\sin\delta_{CP}$ is expected to be small, and the phase is close 
to $\pi$ or zero. This agrees with the result of a  
global fit in \cite{global}: 
\be
\delta_{CP} = \left(1.39^{+0.33}_{-0.27} \right) \pi~~ ({\rm NH}), ~~~~
\left(1.35^{+0.24}_{-0.39} \right)\pi ~~ ({\rm IH}),   
\nonumber
\ee
although statistical significance of this indication is low.  
At a $2\sigma$-level, $\delta_{CP}$ is also consistent 
with zero because of a second local minimum at that value (in both hierarchies). 
The value $\pi/2$, however, is disfavored in both cases. 

Observation of $\delta_{CP} \sim \pi$ would be some indication 
of the CKM scenario or L-R scenario with phase factorization.

\item Observation of a large 
value, $\delta_{CP}\gg\lambda^2$, in experiments will rule out these scenarios
and imply that either there are other sources of CP violation besides the CKM-like phase 
in $U_L$ or that the considered framework (canonical seesaw) is invalid, 
e.g., Dirac mass matrices are non-hierarchical, or seesaw type I is not the mechanism
for generating neutrino masses.

\item In our notation, the effective Majorana mass of the electron neutrino is 
\be
m_{ee}= \left|\sum_i m_i e^{2i \beta_i}U_{ei}^2 \right|, 
\nonumber
\ee  
which, for inverted mass ordering in the limit of hierarchical masses, is mainly 
sensitive to $\beta_1 - \beta_2$.  
Since  $\beta_1 - \beta_2 = {\cal O}(\lambda^3)$,  no cancellation of 
contributions to $m_{ee}$ from the first two mass eigenstates is expected 
and $m_{ee}$ is expected  to be relatively large. For normal  ordering $m_{ee}$ depends mainly on  
the combination $\delta_{CP} + \beta_{2}$. 
Measuring the Majorana phases (or their differences) 
will be challenging for scenarios described above.
 
\item Future precise measurements of the phases may allow to disentangle the 
possibilities: CP in the left rotations only and L-R symmetric case. In the former, 
one expects $\sin\delta_{CP}\gg\beta_{1,2}$, whereas the latter predicts all three 
phases to be equal in the specific case of factorization.

\item If the baryon asymmetry of the Universe is generated via leptogenesis
(decays of the RH neutrinos in our case), this
imposes certain restrictions on structure of the RH sector of seesaw; see, e.g., \cite{lept}
and \cite{leptrew} for recent reviews.
In particular, successful leptogenesis gives the  bounds on  mass of the lightest  RH  neutrino
(in most of the cases we require a strongly hierarchical spectrum) and on combinations
$$
\frac{1}{[U_M^T U_R (m_D^{diag})^2 U_R^{\dagger}U_M^*]_{ii}}
{\rm Im} \left\{ [U_M^T U_R (m_D^{diag})^2 U_R^{\dagger}U_M^*]_{ij} [U_L m_D^{diag} U_R^{\dagger}U_M^*]_{\alpha i}
[U_L m_D^{diag} U_R^{\dagger}U_M^*]^*_{\alpha j}\right\}\,,
$$
where $\alpha = e, \mu, \tau$ is the flavor index and $i, j$ are indices of the RH neutrino mass eigenstates.
The combinations  determine the lepton asymmetries in the lepton channel $\alpha$.
In the case of unflavored leptogenesis a summation over $\alpha$ proceeds, and the dependence on $U_L$ disappears.
So, leptogenesis would require complex phases in $U_R$ and/or  $U_M$. This is not necessary in the flavored case~\cite{lept}.

\end{enumerate}

\section{Conclusions}
\label{sec:conclude}

We have studied the Dirac and Majorana CP violating phases 
in context of the seesaw type I mechanism with similar 
Dirac mass matrices for quarks and leptons. 
In this  case  a relationship $U_{PMNS} = U_L U_X$  is realized with 
$U_{L} \sim V_{CKM}^{\dagger }$.   
We formulated the standard parametrization conditions 
for the mixing matrix to obtain simultaneously both the  Dirac and Majorana CP phases.  
 Possible connections of the Dirac 
CP violating phases in the quark and lepton sectors have been explored. 

The main results that we obtained are:  

\begin{enumerate}

\item  

If the Dirac CP phase in $U_L$ is the only 
 source of CP violation (which is similar to what happens in quark sector  
with  Kobayashi-Maskawa mechanism), 
 and there is no CP violation in the RH sector, 
 the leptonic CP violation is very small  
 $\sin \delta_{CP} = {\cal O}(\lambda^2)$. 
The phase itself is either close to zero or to 
 $\pi$ with the deviation of the order of $\lambda^2$. 
The Majorana phases are expected to be even smaller: 
$\beta_1 \approx \beta_2 = {\cal O}(\lambda^3)$.   

\item 
If the Dirac mass matrices are symmetric 
so that $U_L = U_R \sim V_{CKM} (\delta_q)$ and the Majorana mass matrix of the 
RH fields is real in the L-R symmetric basis, 
$\delta_{CP}$ is in general enhanced by the seesaw mechanism. 
Furthermore, the Dirac masses of the neutrinos are constrained to be not strongly hierarchical.
to reproduce the correct mixing. 

\item
The seesaw  enhancement of phase is absent  if $M_R$ has a specific form  
that leads to the phase factorization in $M_X$. In this case, 
$\beta_1 \approx \beta_2  =  {\cal O}(\lambda^2)$ 
$\alpha_\tau = {\cal O}(\lambda^2)$ and $\sin \delta_{CP} = {\cal O}(\lambda^2)$. 
In particular case $\alpha_\tau = \phi_{td}$   
three phases  
are equal and small $\beta_1 \approx \beta_2 \approx \sin \delta_{CP}  
= {\cal O}(\lambda^2)$. 
Thus, the presence of the CP violation in the RH sector in the factorization case 
enhances the Majorana phases, but keeps the Dirac phase 
at the same order for this scenario. 

\item

Generic CP violation in the RH sector 
can lead to arbitrary and independent values 
of all three phases for arbitrary hierarchy of the eigenvalues of $m_D$. 
We identify that the observable CP phase depends 
mainly on $\alpha_e$, $\alpha_\mu$, and $\delta_X$, if it is measured to be large. 

\end{enumerate}

The formalism developed here allows to explore implications 
of measurements of the CP phases for the RH sector. For example, 
if a large CP phase is observed, the observable CP phases will mainly depend on three 
unknown phases in the RH sector : $\alpha_e$, $\alpha_\mu$, and  $\delta_X$. Thus, 
determination of $\delta_{CP}$ and the Majorana phase may provide information on these 
parameters.
 
We may also get some direct hints about the flavor symmetry 
and quark-lepton unification, if special values of the CP violating phases are observed 
or if certain correlations between them are seen.  
Coming back to the initial question about  the quark and leptonic CP phases, 
even in the context in which quarks and leptons are maximally related 
(quark-lepton symmetry, seesaw type I) one cannot expect equality of the 
quark and lepton Dirac phases. The phases are related but, generically,  
strongly different. The difference can be related to different  mixing angles 
(especially 1-3 mixing angle) and to seesaw mechanism itself. 

Some results of this paper can be modified by the RGE effects. 
Since the light neutrino spectra  we have considered are hierarchical, the 
renormalization correction are small and they will not affect our conclusions. 
The threshold effects due to possible large hierarchy of masses of the RH neutrinos
are important when implications for $M_R$ are considered but this is beyond the scope of this paper.

\section*{Acknowledgment}

One of us A.Y.S. would like to thank W. Rodejohann and  
T. Schwetz-Mangold for discussions and M. Rebelo for useful communications.

\section*{Appendix A: Solution of the Standard  Parametrization Conditions}

In this appendix we provide details of computations of the CP phases 
using the standard parametrization conditions. 

\subsection*{CP violation from CKM only}
 
Using explicit expressions for $U_{e1}$ and $U_{e2}$
in (\ref{eq:uli}), we obtain from the 
conditions ${\rm Arg\,} \{U_{e1}\} = {\rm Arg\,} \{ U_{e2}\} = 0$ 
that 
\be
\beta_1 = \gamma_e + \xi_{e1}, ~~~\beta_2 = \gamma_e + \xi_{e2}, 
\label{beta12}
\ee
where $\xi_{ei}$ are given by 
\bea 
\xi_{e1} & = & - \frac{|V_{td}|X_{\tau 1}}{V_{ud}X_{e1}}\sin\phi_{td} + {\cal O}(\lambda^4)\,,
\nonumber\\
\xi_{e2} & = & - \frac{|V_{td}|X_{\tau 2}}{V_{ud}X_{e2}}\sin\phi_{td}+{\cal O}(\lambda^4)\,. 
\label{xiphase}
\eea 
We see that $\xi_{ei} = {\cal O} (\lambda^3)$, which means $\sin(\eta_i +\gamma_e)$ 
is of the order $\lambda^3$.
The reason behind this is that the CP violation originates from the 
Kobayashi-Maskawa phase associated with the element suppressed by 
$\lambda^3$, while one of real terms in (\ref{eq:uli})
is always of the order 1. Similarly, using  (\ref{eq:uli}), with $\alpha_l = \delta_X = 0$,
and the conditions ${\rm Arg\,} \{U_{\mu 3}\} = {\rm Arg\,} \{U_{\tau 3}\} = 0$,
we find 
\bea
\gamma_\mu&=&{\cal O}(\lambda^4)\nonumber\label{eq:gmu}\\
\gamma_\tau&=& 
\frac{|V_{ub}|X_{e 3}}{V_{tb}X_{\tau 3}}\sin\phi_{ub}+{\cal O}(\lambda^4)\,.
\nonumber
\label{eq:gtau}
\eea
As we will show,  $X_{e3}\leq{\cal O}(\lambda)$, 
so that $\sin\gamma_\tau$ is also at most order $\lambda^4$.
 
Neglecting phases $\gamma_\mu$ and $\gamma_\tau$, 
in the lowest order the 5th condition (\ref{eq:st-cond2}) becomes
\be
X_{e1} X_{\mu 2}  \sin \beta_1 = 
X_{e2} X_{\mu 1}  \sin \beta_2. 
\nonumber
\label{fifth2}
\ee
Then it follows using (\ref{beta12}, \ref{xiphase}) that
\be
\gamma_e = \frac{r \xi_{e1} - \xi_{e2}}{1 - r} , ~~~~~
r \equiv \frac{X_{e2} X_{\mu 1}}{X_{e1} X_{\mu 2}}, 
\label{eq:beta12a}
\ee
and explicitly 
\be
\gamma_e =  \frac{|V_{td}|(X_{e1}^2X_{\mu 2}X_{\tau 2} 
- X_{e2}^2X_{\mu 1}X_{\tau 1})\sin\phi_{td}}{V_{ud} X_{e1} X_{e2} X_{\tau 3}} 
+ {\cal O}(\lambda^4)\,   
\label{eq:ge}
\ee
which shows that $\sin\gamma_e = {\cal O}(\lambda^3)$. 
For Majorana phases we have 
\be
\beta_1 =  \frac{\xi_{e1} - \xi_{e2}}{1 - r} ~~~~ \beta_2  = r \beta_1.
\label{eq:beta12b}
\ee
\\

\subsection*{Left-Right symmetry with factorization} 

Let us consider  $\delta_{CP}$ in the presence of  $\alpha_\tau \neq 0$. 
From (\ref{eq:Ue3}) we have  $s_{13} = |U_{e3}| \approx |A^\prime|$,  where 
\be
A^\prime  \equiv  V_{ud} X_{e3} - |V_{cd}| X_{\mu3} + |V_{td}| X_{\tau 3} \approx A. 
\ee
We can then rewrite Eq. (\ref{eq:cpphase})  
neglecting $\phi_{cd}$  as 
\begin{equation}
\sin \delta_{CP} = 
- {\rm sign} \{ A \}\sin \gamma_e - \frac{1}{s_{13}} \sin (\alpha_\tau - \phi_{td}) V_{td} X_{\tau 3}. 
\label{eq:cpphase2}
\end{equation}
Nonzero $\alpha_\tau$ modifies the phases in (\ref{xiphase}), 
\be
\xi_{e1}  =  \frac{|V_{td}|X_{\tau 1}}{V_{ud}X_{e1}}\sin (\alpha_\tau - \phi_{td})\,, ~~~~
\xi_{e2} =  \frac{|V_{td}|X_{\tau 2}}{V_{ud}X_{e2}}\sin (\alpha_\tau - \phi_{td}).  
\nonumber
\label{xiphase-2}
\ee 
So,  with high accuracy $\beta_1 = \beta_2  \equiv \beta$,  
and consequently,  $\gamma_e - \beta = O(\lambda^3)$. 

From the conditions ${\rm Im}\,U_{\mu 3} = 0$ 
we obtain 
\be
\gamma_\mu + \xi_{\mu 3} = 0, ~~~~ \xi_{\mu 3} = - \frac{V_{ts} X_{\tau 3}}{|V_{cs}| X_{\mu3}} 
\sin \alpha_\tau ,     
\label{eq:gammamu}
\ee
so that  $\xi_{\mu 3} = {\cal O} (\lambda^2)$.  
The equality  ${\rm Im}\,U_{\tau 3} = 0$ gives 
$
\gamma_\tau + \alpha_\tau = |V_{cb}| X_{\mu 3} / (|V_{tb}| X_{\tau 3}) \sin\alpha_\tau. 
$
The 5th SP condition (\ref{eq:st-cond2}), gives at the leading order
\be
\big[|V_{cs}| X_{\mu 1} \sin (\gamma_\mu - \beta) + 
|V_{ts}| X_{\tau 1} \sin \alpha_\tau \big] X_{e2} =  
\big[ |V_{cs}|X_{\mu 2}\sin (\gamma_\mu - \beta) + 
|V_{ts}| X_{\tau 2}\sin \alpha_\tau \big] X_{e1},  
\nonumber
\label{eq:stpar2}
\ee
which leads to  
\be
\sin (\gamma_\mu - \beta) = \sin \alpha_\tau \frac{|V_{ts}| X_{\mu 3}}{|V_{cs}| X_{\tau 3}}. 
\nonumber
\ee
Using expression  for $\gamma_\mu$ from (\ref{eq:gammamu}) 
we obtain 
\be
\sin \beta = \sin \gamma_e = \sin \alpha_\tau 
\frac{|V_{ts}|}{|V_{cs}|}~ \frac{ X_{\tau 3}^2 - X_{\mu3}^2}{X_{\tau 3} X_{\mu3}}
= \sin \alpha_\tau \frac{|V_{ts}|}{|V_{cs}|} \frac{2 \cos 2\theta_{23}}{\sin 2\theta_{23}}. 
\label{sinbg}
\ee
Thus,  $\beta = {\cal O} (\lambda^2)$,  
and consequently, $\gamma_e = {\cal O} (\lambda^2)$ or smaller. 
Inserting $\sin \gamma_e$ from (\ref{sinbg}) into (\ref{eq:cpphase2})
we obtain  
\be
\sin \delta_{CP} =  - {\rm sign} \{ A \} \sin \phi_{td} 
\frac{|V_{ts}|}{|V_{cs}|} \frac{2 \cos 2\theta_{23}}{\sin 2\theta_{23}}
- \frac{1}{s_{13}} \sin (\alpha_\tau - \phi_{td}) V_{td} X_{\tau 3}.
\label{cpcp}
\ee
According to (\ref{cpcp}) effect of non-zero $\alpha_\tau$, 
{i.e.} from the RH sector, is of the same order as the result for the CKM phase only. 
If $X_{\mu3} = X_{\tau 3}$, that is the 2 - 3 mixing in $U_X$ is maximal  
$\beta =  0$, but 
$$
\delta_{CP} = - \frac{|V_{td}|}{\sqrt{2} s_{13}} \sin (\alpha_\tau - \phi_{td}) V_{td}.  
$$

\section*{Appendix B: Expressions for elements of the $U_X$ matrix} 

For a real $U_X$, using the relations (\ref{eq:uli}) we obtain at the lowest order
\bea
X_{e1} &=& c_{12}/V_{ud}  + {\cal O}(\lambda)\nonumber\label{eq:xspl1}\,,
\\
X_{e2} &=& s_{12}/V_{ud} + {\cal O}(\lambda)\nonumber\label{eq:xspl2}\,,
\\
X_{\mu3} &=& s_{23}/{|V_{cs}|} + {\cal O}(\lambda^2)\nonumber\label{eq:xspl4}\,,
\\
X_{\tau3} &=& c_{23}/V_{tb} + {\cal O}(\lambda^2)\label{eq:xspl5}\,.
\eea

Using smallness of $X_{e3}$ the elements $X_{\mu 1}$, 
$X_{\mu 2}$, $X_{\tau 1}$, and $X_{\tau 2}$ are expressed in terms 
of the above 4  elements and $X_{e3}$ as 
\bea
X_{\mu 1} &=& -X_{e2}X_{\tau3} - X_{e1}X_{\mu3}X_{e3} + {\cal O}(\lambda^2)\,,\nonumber\label{eq:xspl6}\\
X_{\mu 2} &=& X_{e1}X_{\tau3} - X_{e2}X_{\mu3}X_{e3}+ {\cal O}(\lambda^2)\,,\nonumber\label{eq:xspl7}\\
X_{\tau 1} &=& X_{e2}X_{\mu3} - X_{e1}X_{\tau3}X_{e3}+ {\cal O}(\lambda^2)\,,\nonumber\label{eq:xspl8}\\
X_{\tau 2} &=& -X_{e1}X_{\mu3} - X_{e2}X_{\tau3}X_{e3}+ {\cal O}(\lambda^2)\label{eq:xspl9}\, .  
\eea

\end{document}